\documentclass[preprint2]{aastex}% preprint2 produces a double-column, single-spaced document:
\usepackage{natbib}
\shorttitle{Radial Stellar Pulsation and Three-dimensional Convection. III.}
\shortauthors{Geroux and Deupree}

\begin{document}

\title{RADIAL STELLAR PULSATION AND THREE-DIMENSIONAL CONVECTION. III. COMPARISON OF TWO-DIMENSIONAL AND THREE-DIMENSIONAL CONVECTION EFFECTS ON RADIAL PULSATION}

\author{Christopher M. Geroux\altaffilmark{1} and Robert G. Deupree}
\affil{Institute for Computational Astrophysics and Department of Astronomy and Physics, Saint Mary's University, Halifax, NS B3H 3C3 Canada}
\email{geroux@astro.ex.ac.uk}
\altaffiltext{1}{Now at Physics and Astronomy, University of Exeter, Stocker Road, Exeter, UK EX4 4QL}

\begin{abstract}
We have developed a multidimensional radiation hydrodynamics code to simulate the interaction of radial stellar pulsation and convection for full amplitude pulsating models. Convection is computed using large eddy simulations. Here we perform three-dimensional simulations of RR Lyrae stars for comparison with previously reported two-dimensional simulations. We find that the time dependent behavior of the peak convective flux on pulsation phase is very similar in both the two-dimensional and three-dimensional calculations. The growth rates of the pulsation in the two-dimensional calculations are about 0.1\% higher than in the three-dimensional calculations.The amplitude of the light curve for a 6500~K RR Lyrae model is essentially the same for our 2D and 3D calculations, as is the rising light curve. There are differences in slope at various times during falling light.
\end{abstract}

\keywords{convection --- hydrodynamics --- methods: numerical --- stars: oscillations --- stars: variables: general --- stars: variables: RR Lyrae}

\section{INTRODUCTION}
RR Lyrae and classical Cepheid variables have long played an important role as standard candles in the development of our understanding of the structure and evolution of our galaxy and of nearby galaxies. Their importance has led to a long history of trying to model the pulsation of these variables \citep[e.g.][]{Christy-1964,Christy-1966b,Cox-1966a,Stellingwerf-1975,Bono-1994} with one-dimensional hydrodynamic simulations. These were successful in computing a number of full amplitude light curves that agreed in some detail with those observed, at least as long as the models were not chosen too close to the red edge of the instability strip. It was speculated early on that convection in the ionization regions for models near the red edge would be important \citep{Christy-1966a,Cox-1966b}, but neither models in which mixing length convection was “frozen in” at static model levels \citep[e.g.][]{Tuggle-1973} nor models in which convection instantaneously adjusted to the static flux for the current state variables \citep{Cox-1966b} predicted a return to stability at the red edge. This led to the development of more sophisticated time dependent mixing length approaches \citep[e.g.][]{Stellingwerf-1982a,Stellingwerf-1982b,Stellingwerf-1984a,Stellingwerf-1984b,Stellingwerf-1984c,Kuhfuss-1986,Xiong-1989}, and  one dimensional hydrodynamic simulations using convective models such as these were able to compute a red edge \citep[e.g.][]{Bono-1994,Gehmeyr-1992a,Gehmeyr-1992b,Gehmeyr-1993}. Further calculations \citep[e.g.][]{Bono-1997a,Bono-1997b,Marconi-2003,Marconi-2007} were able to produce full amplitude light curves of RR Lyrae variables which somewhat resemble what is observed, although the agreement between the observed and computed light curves for low amplitude RR Lyrae variables near the red edge remains relatively poor \citep{Marconi-2007}. The general conclusion appears to be that the treatment of convection in pulsating stars remains unsatisfactory \citep[e.g.][]{Buchler-2009,Marconi-2009} as evidenced  by the relatively poor agreement between the observed and computed light curves near the red edge of the RR Lyrae gap.

The problems with the local mixing length approach, including the necessity of assuming values of free parameters which can significantly change the solution, led \cite{Deupree-1977a} to approach the problem of the interaction between convection and radial pulsation in RR Lyrae variables in a different way. He performed two-dimensional (2D) hydrodynamic simulations following the largest eddy developed in two dimensions by the convective instability in the hydrogen ionization zone. This allowed him to determine the red edge location \citep{Deupree-1977b}  and to show that the first overtone red edge was to the blue of the fundamental red edge \citep{Deupree-1977c}. The reason for the red edge is that convection essentially allows energy transport out of the hydrogen ionization region when pulsational instability needs to store it and stops transporting energy near maximum velocity when it needs to be released to drive the pulsation. However, he was not able to compute full amplitude models because the algorithm he used to determine the radial flow of his mesh (by forcing the mesh to move at the horizontal average radial velocity at each radial mesh point) could not keep the very narrow hydrogen ionization zone resolved in the mesh for more than about twenty periods. There have been some other 2D calculations undertaken recently to study the interaction of convection and pulsation \citep{Mundprecht-2012,Gastine-2011}, but these have not yet led to a comparison of full amplitude solutions with observations.

The problem with the mesh propagation in multi-dimensional calculations has been successfully solved by \cite{Geroux-2011}, who devised a radial mesh flow algorithm in which the mass in a given spherical shell does not vary during the course of the calculation. This does not mean the calculation is Lagrangian; it merely means that there is no net mass flow out of a spherical shell. This is a comparatively simple version of techniques where the computational mesh is allowed to move according to certain rules \citep[e.g.][]{Gehmeyr-1992a,Dorfi-1991,Feuchtinger-1996}. The horizontal motion is determined by the conservation laws, and mass can flow into and out of a spherical shell; there just cannot be any net mass flow out of the shell. This has allowed the calculation of full amplitude pulsation models in 2D \citep{Geroux-2013b}. The primary results of these calculations are that the light curves resemble those produced by one-dimensional codes for models not close to the red edge, although the amplitudes tend to be somewhat lower, while models near the red edge agree much better with the observed light curves than do those of \cite{Marconi-2007}. These 2D calculations did not produce a red edge, however, because the convective region wanted to grow into the regions well below the ionization regions as the models became cooler and the pulsation amplitude became larger. This significantly changed the structure and potential energy and thus thermal energy content in regions of the model just interior to the ionization regions. The thermal relaxation time of these deeper regions is sufficiently long that it is impractical to follow the evolution to a full amplitude solution corresponding to the newer structure with an explicit hydrodynamic calculation.

We appreciate that convection is not a 2D phenomenon. The argument made by \cite{Deupree-1977a} and by \cite{Geroux-2013b} is that the time dependence of convection is possibly more important in determining the pulsation behavior than the details of the convective flow. This is clearly an assumption, and we are now in a position to examine this by the computation of three-dimensional (3D) convection and pulsation for comparison with the \cite{Geroux-2013b} 2D results. The physics and model input in this paper are the same as for the 2D calculations. The calculations are made with the OPAL opacities \citep{Iglesias-1996} in conjunction with the low temperature \citep{Alexander-1994} opacities. Radiation is treated with the diffusion approximation everywhere. The OPAL equation of state \citep{Rogers-1996} is used throughout. Convection is treated as a large eddy 2D or 3D flow simulation depending on the calculation, with a subgrid scale eddy viscosity approach to mimic the effects of the small scale convective flow that cannot be resolved in the mesh. The equations and more details are given by \cite{Geroux-2013b}. Each calculation in this paper uses 16 processors.

In this paper we will compare 2D and 3D models both during the pulsational growth for several models and at full amplitude for one specific calculation. As one can imagine, the 3D calculations are quite time consuming, and it will be a few more months before all models are complete to full amplitude. In the next section we compare the 2D and 3D convective flow patterns. In section~3 we examine how the difference between 2D and 3D convection effect the radial pulsation growth rates and how the convective strength depends on pulsation amplitude. In Section~4 we consider the differences in time dependent behavior of full amplitude pulsation with 2D and 3D convection.

\section{CONVECTIVE FLOW PATTERNS}
\label{sec:con-flow-patterns}
We have performed simulations of RR Lyrae pulsation with 3D convection at effective temperatures of 6200, 6300, 6400, 6500, 6700, and 6900 K. The initial parameters of these models match their 2D counterparts presented by \cite{Geroux-2013b}--$L=50 L_\odot$, $M=0.7 M_\odot$,  $X=0.7595$, and $Z=0.0005$. The difference is that these models have the extra dimension for fluid flow. Given the highly turbulent nature of convection in the surface ionization regions of RR Lyrae stars, the convective motion should be 3D. These 3D simulations have the same radial and $\theta$-zoning (140$\times$20) as the 2D calculations but also have 20 $\phi$ zones covering 6$^\circ$, producing a 3D version of a pie slice subtending 36 square degrees. The choice of 6$^\circ$ coverage in each direction comes from relatively short 3D simulations with angular zoning which subtended total angles from $2^\circ\times 2^\circ$ to $10^\circ\times 10^\circ$ with both $\theta$ and $\phi$ stepping simultaneously in increments of $2^\circ$ between the two extremes. These short simulations were for a 5700~K effective temperature model with strong convection and were carried out until convection had finished growing from machine round off errors and at least two additional pulsation cycles had been completed. The $6^\circ\times 6^\circ$ configuration was found to be a good compromise between the inclusion of multiple convective cells and good resolution. The $6^\circ$ simulation was the smallest angular coverage for which we found more than one distinct convective cell. We have performed short 3D simulations with the number of $\theta$ and $\phi$ zones of $5\times 5$ up to zonings of $40\times 40$. Simulations with the largest number of angular zones had very large computational requirements and the amount of computational time required to reach full amplitude would have been prohibitively long. As a compromise we chose 20 $\theta$ and 20 $\phi$ zones, zoning which is still quite computationally demanding (these calculations require several months). It should be emphasized that the calculation time per time step is quite reasonable; however, the number of time steps required to obtain a full amplitude solution is large.

\begin{figure}
\center
\plotone{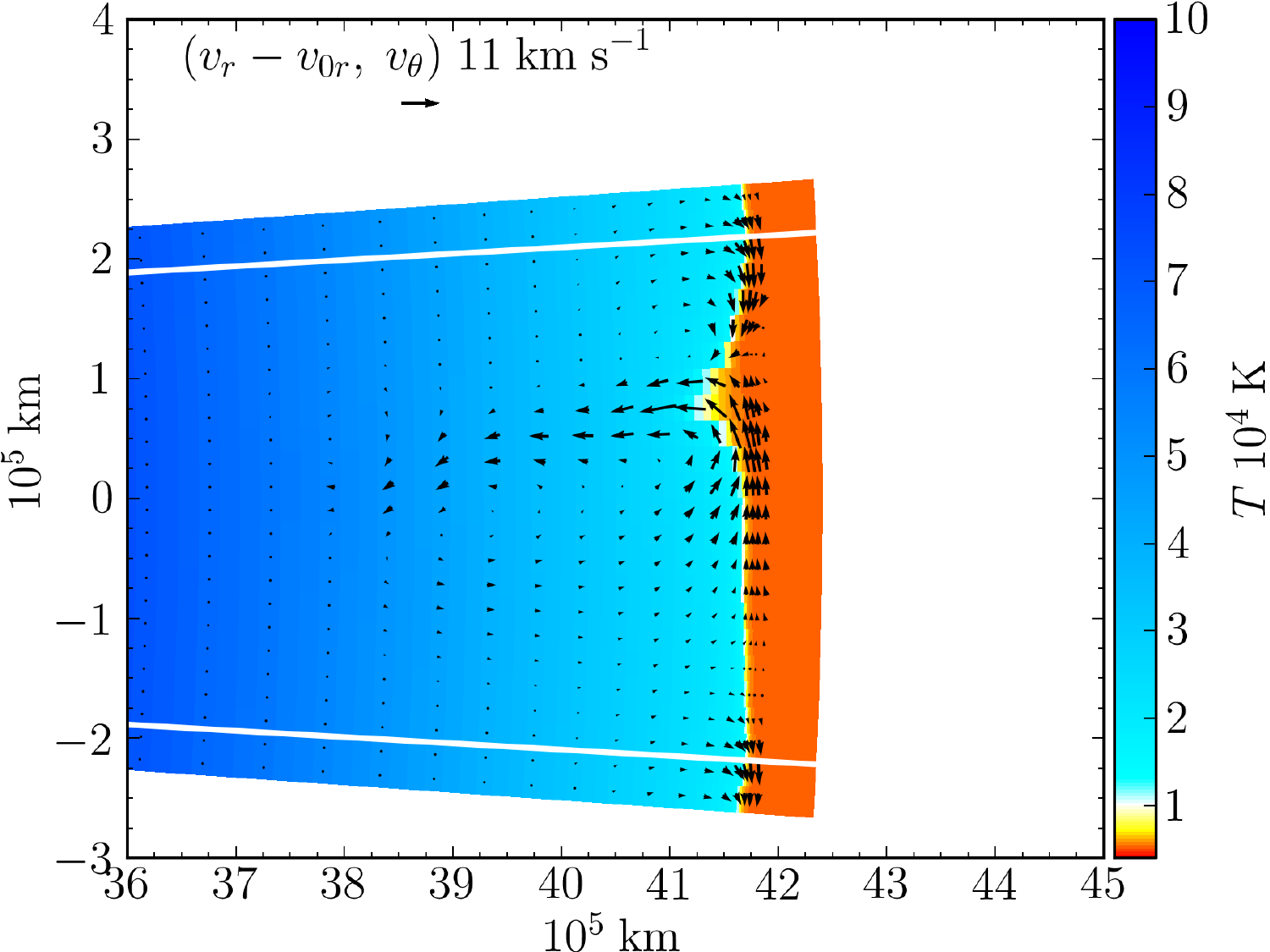}
\caption{Upper 16\% by radius of a 2D simulation of a 6300~K effective temperature RR Lyrae model. Temperature is indicated by the color scale, and the vectors show the direction of the convective flow. Note the relatively narrow, high velocity downward convective flow in comparison to the slower moving wider area upward flow. The white radial lines indicate the horizontal periodic boundaries of the calculation.}
\label{fig:2D-flow-pattern}
\end{figure}

\begin{figure}
\center
\plotone{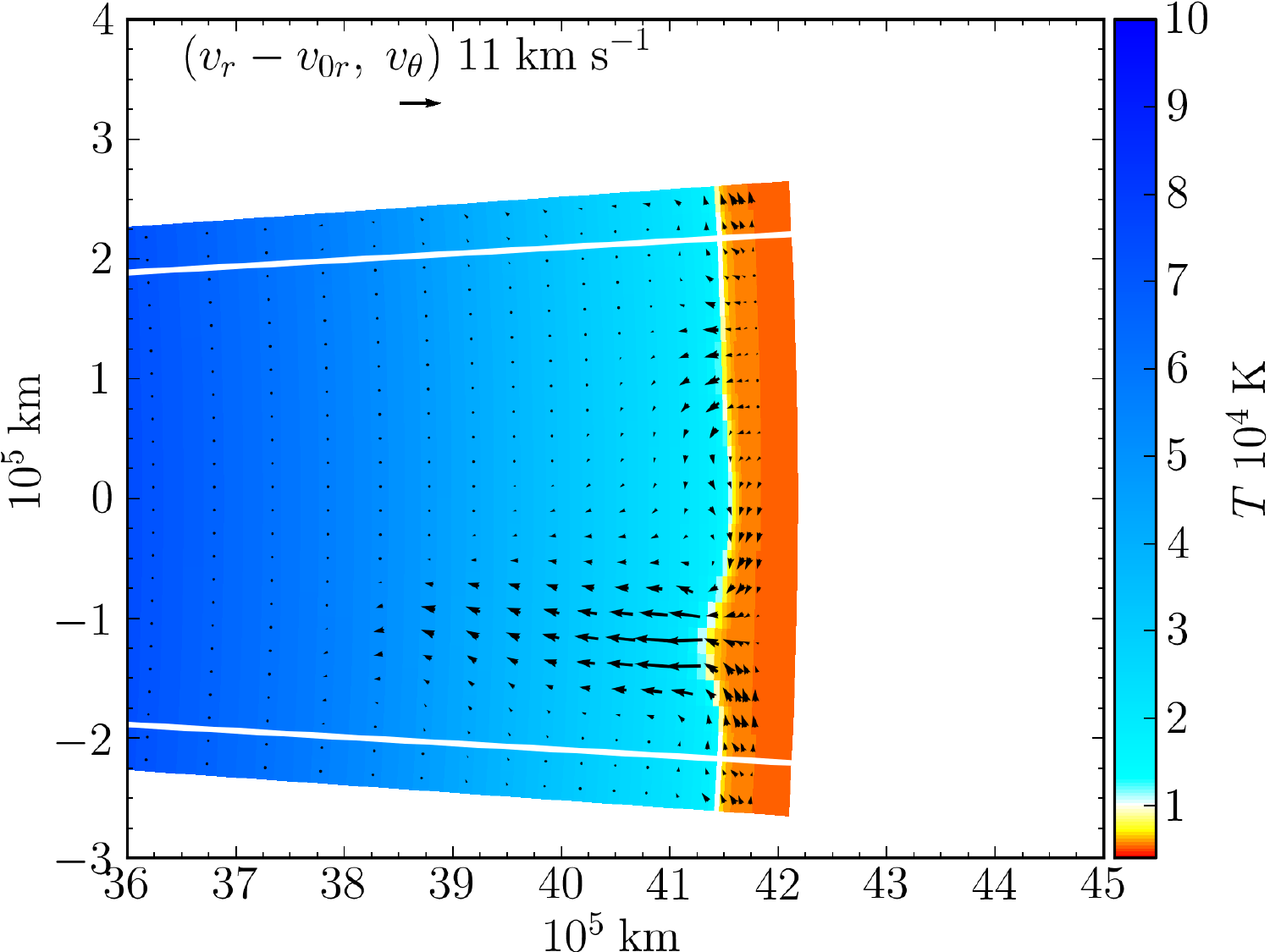}
\caption{Upper 16\% by radius of an $r$--$\theta$ slice through a 3D simulation of a 6300~K effective temperature RR Lyrae model. Temperature is indicated by the color scale, and vectors show the motion in the $r$--$\theta$ plane. Note that, in contrast to Figure~\ref{fig:2D-flow-pattern}, the downward motion covers a wider area and that there is little evidence of upward flow in this particular plane.}
\label{fig:3D-flow-pattern}
\end{figure}

We begin by comparing the flow patterns associated with the convective motion. Figure~\ref{fig:2D-flow-pattern} shows the top 16\% by radius of the 6300~K effective temperature 2D simulation. The color shows the temperature of the material and the vectors show the convective velocity. The white lines show the horizontal periodic boundaries. Figure~\ref{fig:3D-flow-pattern} is similar to Figure~\ref{fig:2D-flow-pattern} but shows a slice through the comparable 3D simulation. The convective flow pattern of the 2D simulation at first glance appears similar to a slice through a comparable 3D simulation. However, there are some differences in the flow pattern. In particular the circular flow pattern clearly visible in the 2D simulation is not as noticeable in the slice through the 3D simulation. This may be a result of the fact that the extra dimension allows some of the return flow to take place in a different plane.

\begin{figure}
\center
\plotone{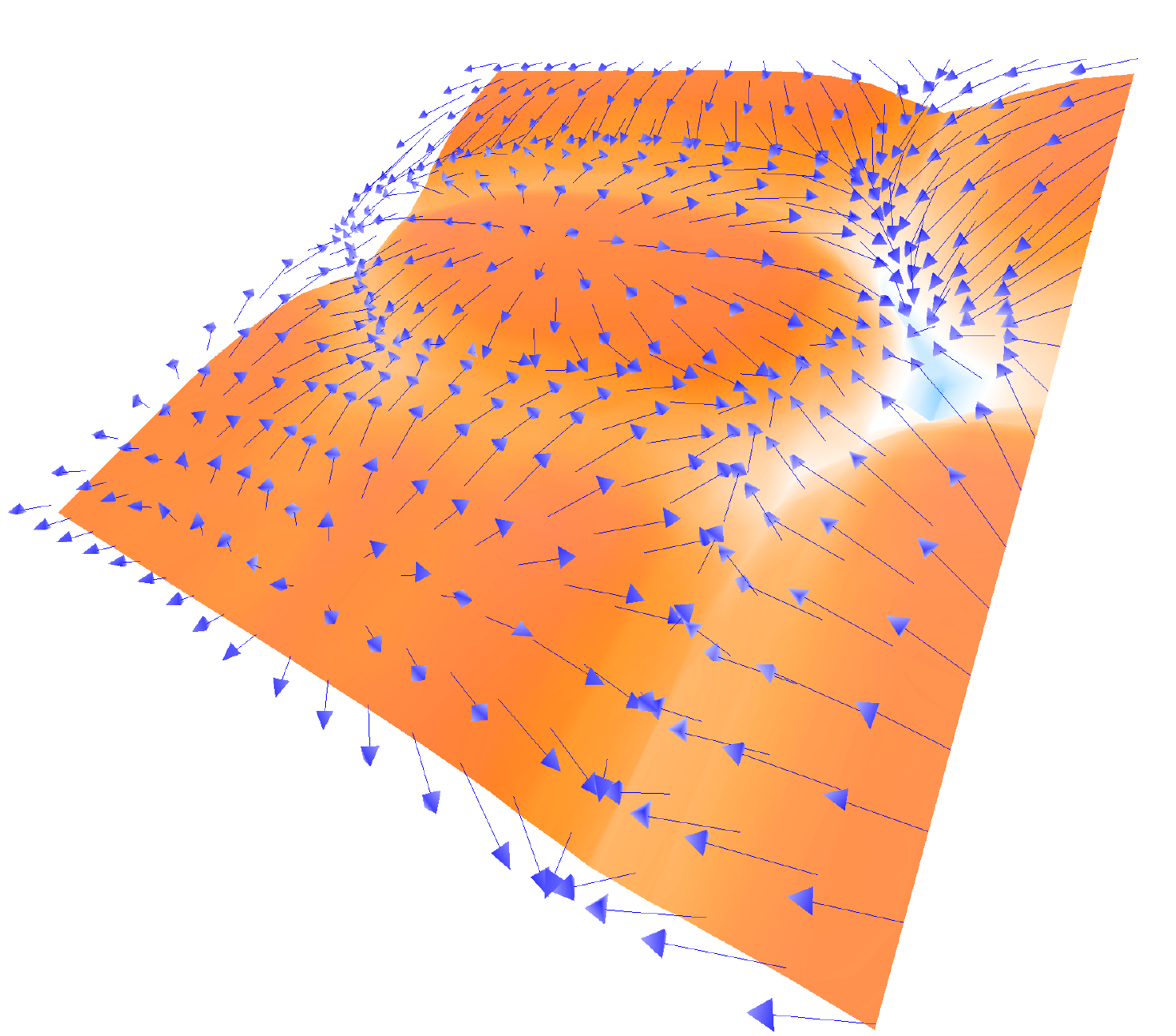}
\caption{Temperature isosurface ($T=10^4$~K) and convective velocity vectors for points on a horizontal plane above the isosurface. The color of the isosurface indicates upward convective motion in red and downward convective motion in blue.  This ``snapshot'' is taken during radial pulsation contraction for a 6300~K effective temperature model.}
\label{fig:temp-iso-6300-con}
\end{figure}

\begin{figure}
\center
\plotone{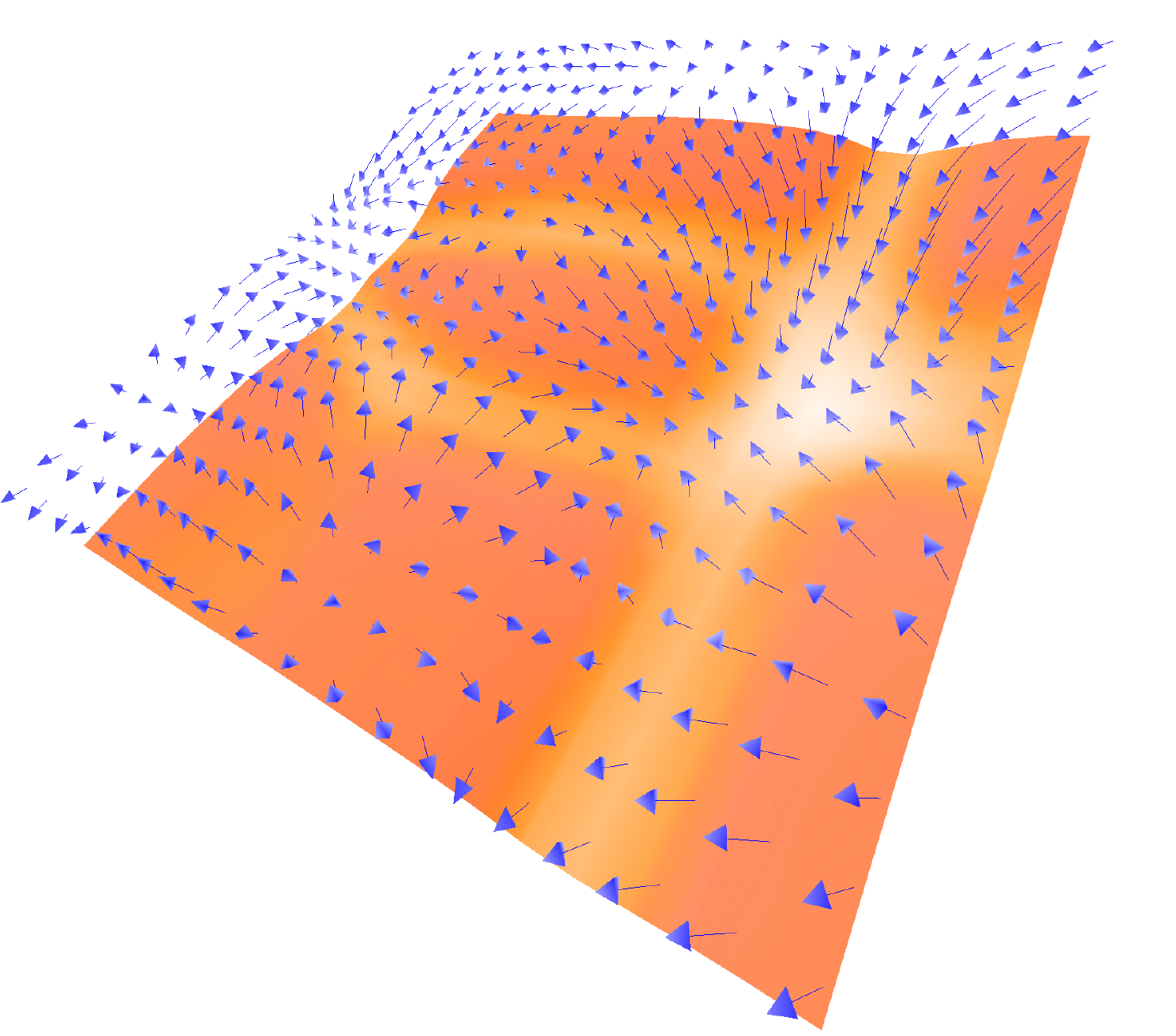}
\caption{Similar to Figure~\ref{fig:temp-iso-6300-con} except during radial pulsation expansion instead of radial pulsation contraction. Notice that the convective velocities are larger during contraction than expansion.
}
\label{fig:temp-iso-6300-exp}
\end{figure}

\begin{figure}
\center
\plotone{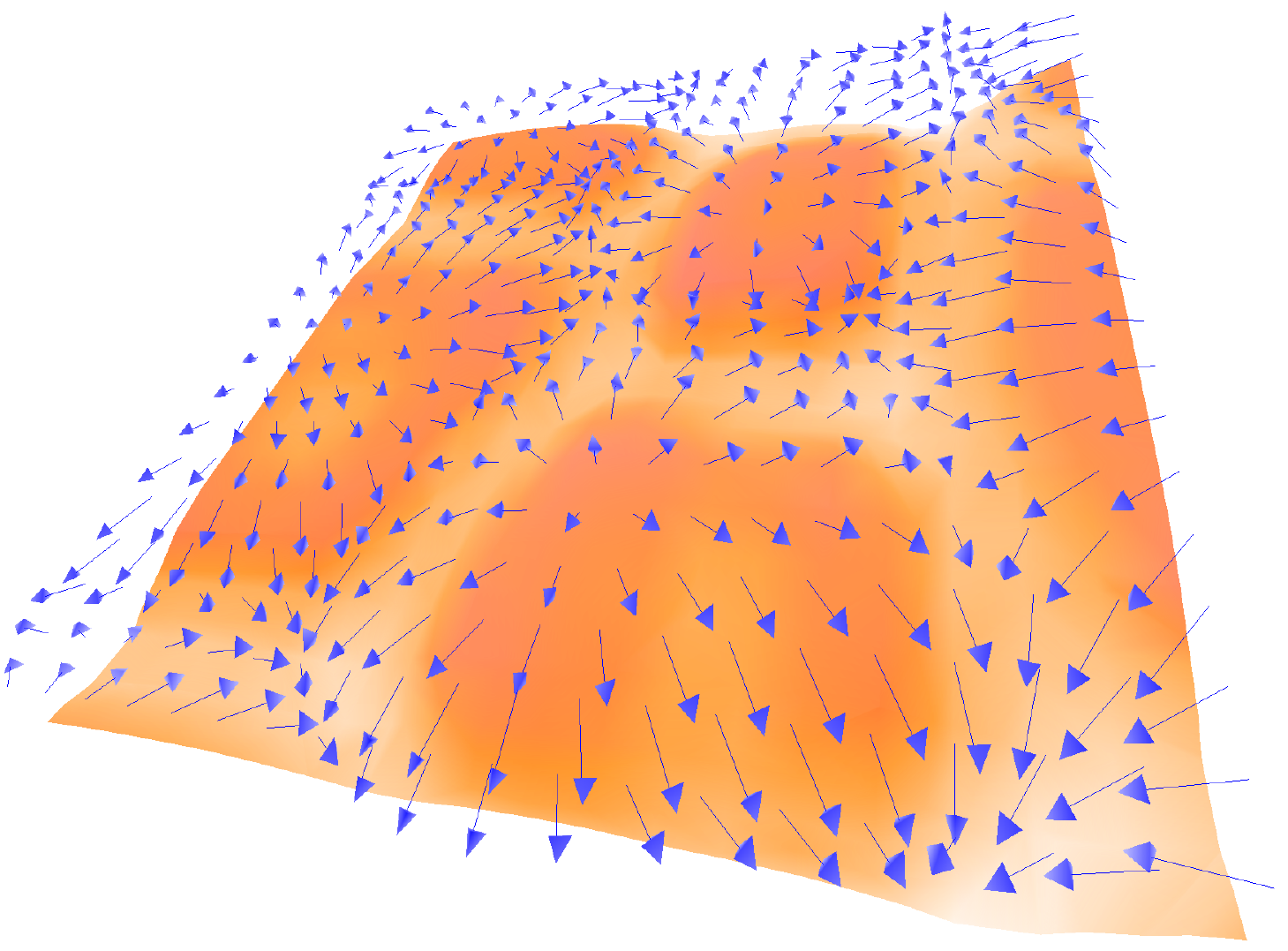}
\caption{Similar to Figure~\ref{fig:temp-iso-6300-con} except for a model with an effective temperature of 6700~K during radial pulsation contraction.
}
\label{fig:temp-iso-6700-con}
\end{figure}

\begin{figure}
\center
\plotone{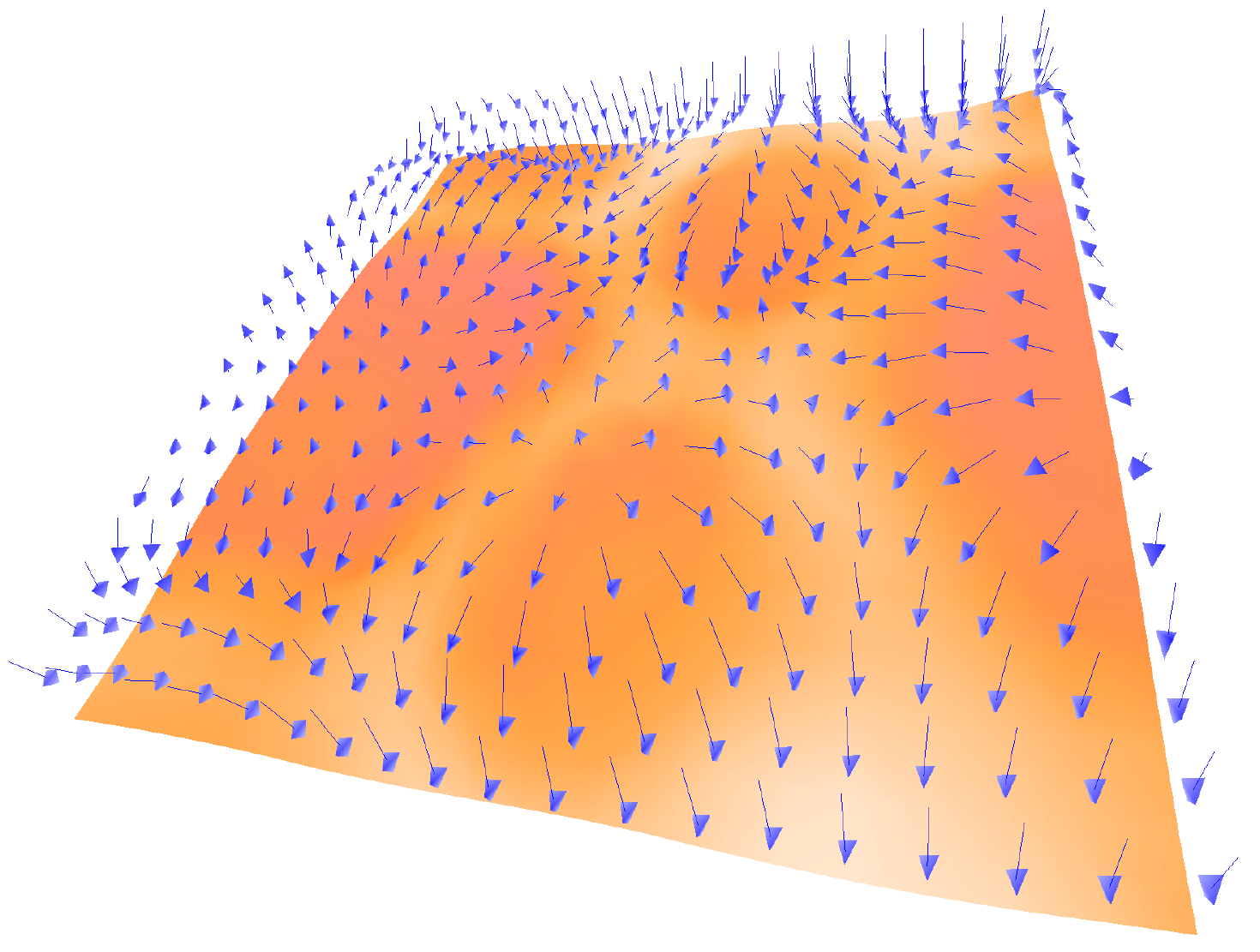}
\caption{Similar to Figure~\ref{fig:temp-iso-6700-con} except during radial pulsation expansion instead of contraction.
}
\label{fig:temp-iso-6700-exp}
\end{figure}

While Figure~\ref{fig:3D-flow-pattern} provides information about how the 3D convective flow pattern behaves in the radial and $\theta$ directions, it is more informative to see the flow pattern in both the horizontal directions. Figures~\ref{fig:temp-iso-6300-con} and \ref{fig:temp-iso-6300-exp} show the temperature isosurface at $10^4$~K spanning the full horizontal extent of the 6300~K effective temperature 3D simulation during the pulsation compression and expansion phases, respectively. One can see that the convection truly is 3D in nature. The reduction in convective strength from compression to expansion is clear, with larger velocity vectors and larger variations in the temperature isosurface showing stronger convective flows during compression and smaller velocity vectors and a flatter temperature isosurface during expansion. Figures~\ref{fig:temp-iso-6700-con} and \ref{fig:temp-iso-6700-exp} are the same as Figures~\ref{fig:temp-iso-6300-con} and \ref{fig:temp-iso-6300-exp} except for the 6700~K effective temperature model. Comparing the figures for the 6300~K effective temperature model to the figures for the 6700~K effective temperature model, it is clear that the change in convective strength from compression to expansion is smaller for the hotter model, although convection remains stronger during contraction than expansion for both models. It is interesting to note that the convective patterns show some similarity to the solar granulation pattern, in that they have large slow moving hot up flows, surrounded by fast narrow cool down flows.

A possible criticism of these calculations is of the small angular extent ($6^\circ \times 6^{\circ}$) only containing two ($T_{\rm eff}=$6300~K model) to four ($T_{\rm eff}=$6700~K model) convective granules as well as the poor resolution ($20 \times 20$ horizontal zones). In an attempt to help validate that our simulations are getting the large scale flows correct we can compare the granule sizes and the up flow filling factor to other 3D simulations of convection in stars. Recently work by \cite{Magic-2013a} has mapped out granule diameters using high resolution 3D atmosphere models and derived a simple relation between the granule diameter and the two parameters $T_{\rm eff}$ and $\log g$. Unfortunately the $T_{\rm eff}$ and $\log g$ of our models (around $\log g=2.8$ and $T_{\rm eff}=6000-7000$~K) have not been modelled by \citeauthor{Magic-2013a}. The closest effective temperature of their models which bracket our gravities is 5500~K. Though the authors caution against extrapolation of their relations it may still give some indication of the size one might expect for granules in our simulations. From \citeauthor{Magic-2013a}'s relations we obtain $10.20$ and $10.18$ for $\log d_{\rm gran}$ (with $d_{\rm gran}$ in cm) for the largest granules in the 6300~K and 6700~K effective temperature models respectively. A simple way to estimate the diameter of the granule in our simulations is to divide the horizontal area of our computational domain by the number of granules and calculating the diameter of the granule from this area. Doing so results in $\log d_{\rm gran}$ of $10.5$ and $10.3$ for the 6300~K and 6700~K models respectively. Our simulations have granules that are slightly larger than predicted from the work by \citeauthor{Magic-2013a} but are the same order of magnitude.

We have also explored the filling factor of up-flows (the fraction of the horizontal area with $v_r-v_0>0$), $f_{\rm up}$,  and found that there is a time dependence on the pulsation phase. The filling factor was measured at the temperature where the temperature gradient is the steepest ($10^4$~K). When the star is fully expanded $f_{\rm up}\approx 0.7$, while when the star is fully contracted $f_{\rm up}\approx 0.4$. The time average of the filling factor over four pulsation phases results in $\langle f_{\rm up}\rangle\approx 0.6$. These values are quite similar for both the 6300~K and 6700~K models examined. The time averaged value of 0.6 is reasonably close to that found by other authors of about $2/3$ \citep{Magic-2013a,Stein-1998}. The similarity of the granule sizes and filling factor to those obtained by higher resolution 3D studies of stellar convection, covering larger horizontal extents provides some confidence that we are calculating the largest scale structures of convection correctly. Despite this order of magnitude agreement, we do not argue that our horizontal zoning is sufficient for our calculations to adequately represent the details of turbulent convection. Given the amount of computer time required for the current 3D calculations, full amplitude 3D calculations with, say, an order of magnitude more zones in each horizontal direction is still some time away.

\section{2D AND 3D DEPENDENCE OF PULSATION GROWTH RATES AND OF CONVECTION ON PULSATION AMPLITUDE}
\label{sec:2D-3D-con-pul-growth}

\begin{figure}
\center
\plotone{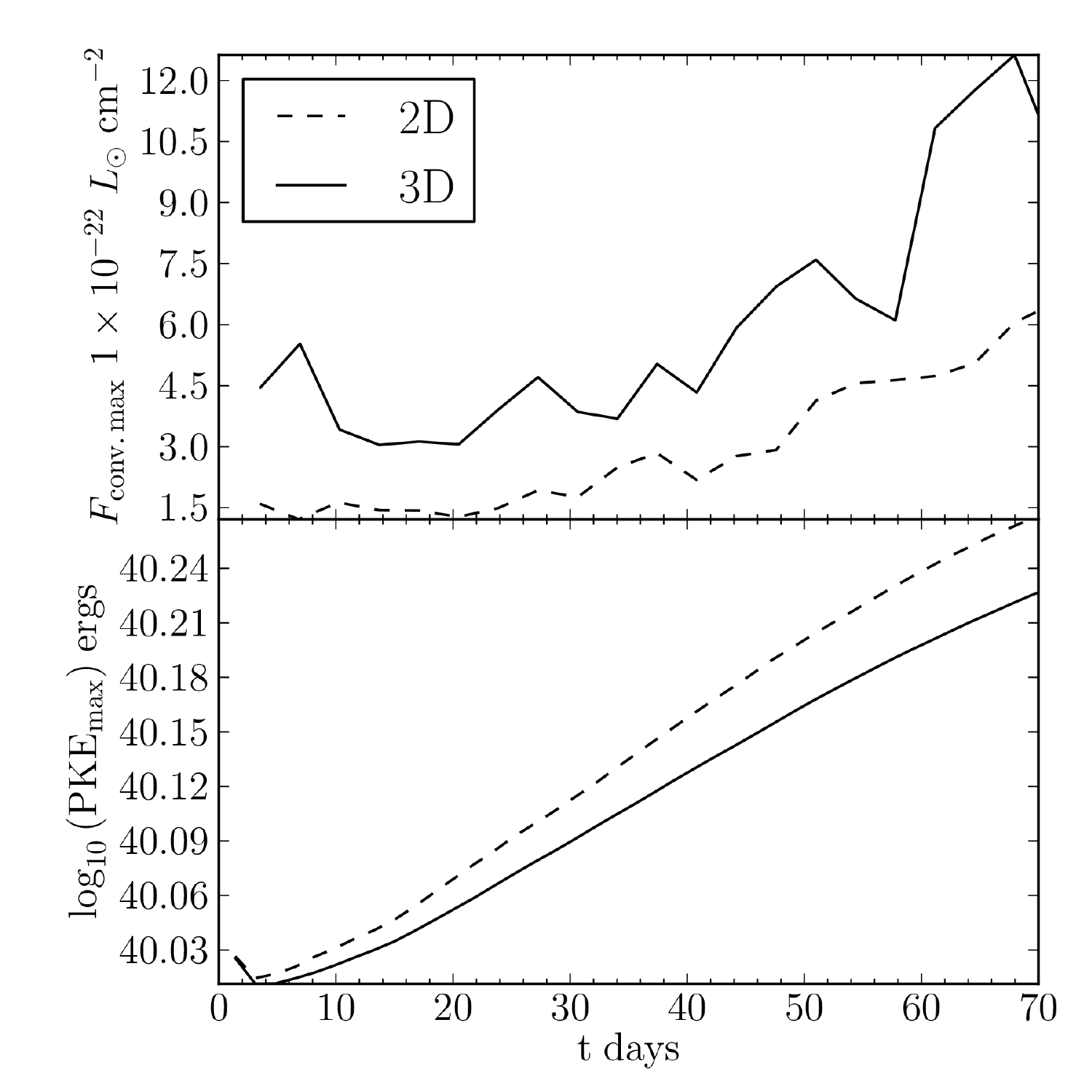}
\caption{Six period average of the peak convective flux during a pulsation period and three period average of the log of the peak kinetic energy per period versus the time since the beginning of the calculation. 2D calculations are denoted by the dashed curves, and 3D by the solid curves. Although the effects of convection on the growth rate are small, they become apparent over many periods. These results are for a 6500~K effective temperature model.
}
\label{fig:PKE-2D-v-3D}
\end{figure}

\cite{Geroux-2013b} showed that the peak convective flux for a pulsating model depended on the pulsation amplitude. Here we wish to see that this remains true in 3D. Figure~\ref{fig:PKE-2D-v-3D} shows how the six period average of peak convective flux per period varies with pulsation amplitude. This average is determined in the following way -- first we find the convective flux for every zone in a given model and select the largest value. We then compare this peak convective flux for all models within a given period and again select the largest value. The six period average is then the average of these single period peak fluxes. The peak convective flux averaged over the six periods clearly increases as the peak kinetic energy of the radial pulsation increases.

Figure~\ref{fig:PKE-2D-v-3D} also shows that the corresponding 3D simulations have a larger peak maximum convective flux for a given peak kinetic energy than for the 2D simulations. This does have an effect on the pulsational growth rates. We determine the growth of the stellar pulsation by using the three period average of the peak kinetic energy per period. The three period average reduces the variation introduced by the first overtone in these fundamental mode calculations. Figure~\ref{fig:PKE-2D-v-3D} indicates that the 3D pulsational growth rate is less than the 2D growth rate. We have calculated the growth rates for the 2D and 3D simulations and find the 2D growth rates are larger than the 3D growth rates by about 0.07–-0.09\% per period, with larger differences for cooler models. This suggests that the relative behavior of convection in 2D and 3D, in terms of its interaction with pulsation, does not vary much across the fundamental mode region of the instability strip. In the calculation of the pulsational kinetic energy it is assumed that the pulsation is given by the radial motion of the coordinate system. Note that this assumption does not affect the numerical simulation; it only affects our interpretation of the results. In fact the pulsational kinetic energy vastly exceeds the convective kinetic energy so that some error in this assumption should not alter our conclusions. We note in this regard that all our calculations are pulsation dominated not convectively dominated, including those near the red edge. This suggests that the density stratification does not strongly limit the pulsation as found in some cases by \cite{Gastine-2011}. The 2D and 3D growth rates are much closer to each other than are the 1D and 2D growth rates given in \cite{Geroux-2013b}, emphasizing that convection in either its 2D or 3D framework helps to slow the pulsational growth.

\section{FULL AMPLITUDE TIME DEPENDENT BEHAVIOR}
\label{sec:phase-dep-con-flux}

\begin{figure}
\center
\plotone{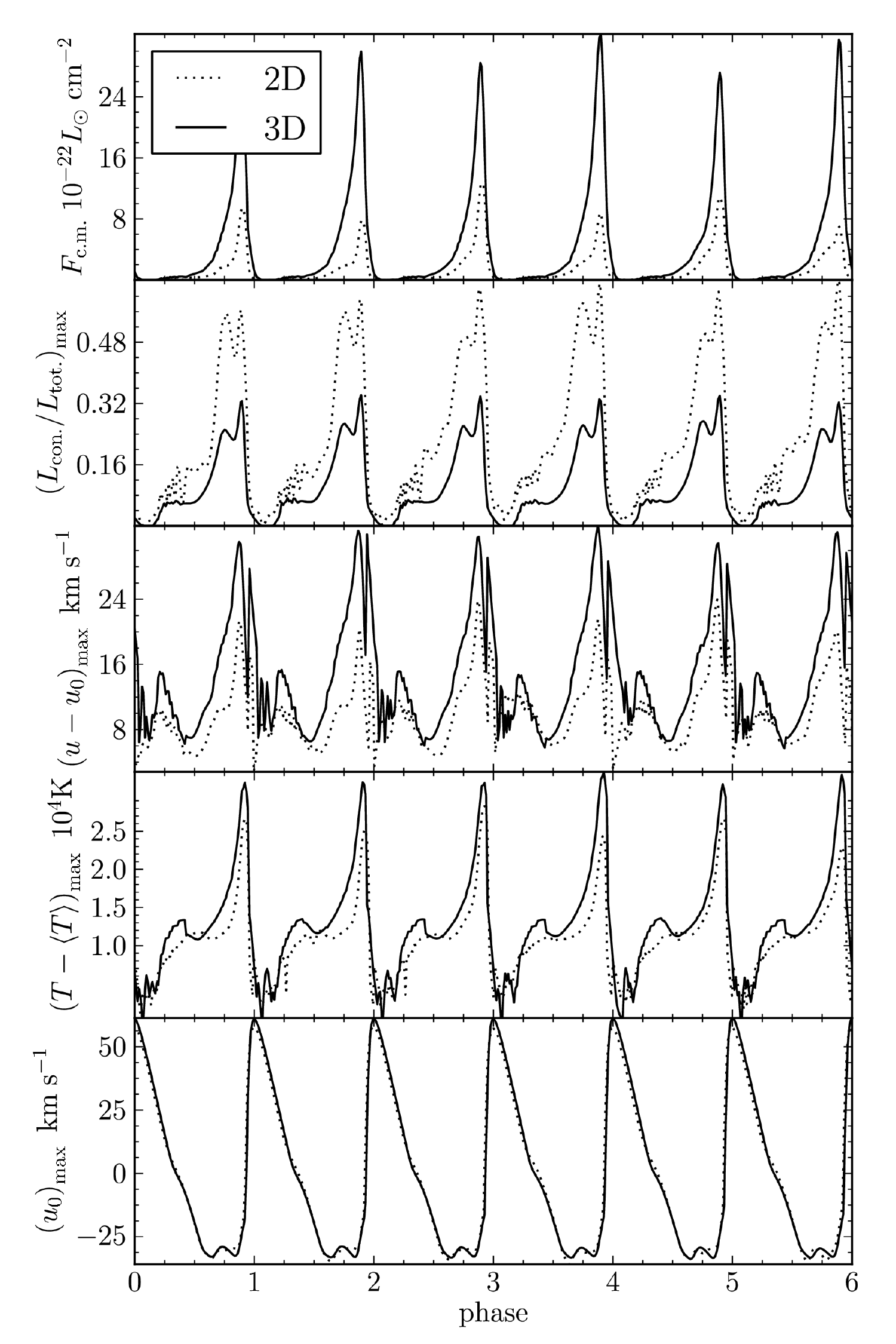}
\caption{From top to bottom: peak convective flux, peak of the ratio of convective luminosity to total luminosity, peak radial convective velocity, maximum variation of the horizontal temperature variation from the horizontally averaged temperature, and surface pulsation velocity for 3D (solid) and 2D (dash) calculations of the 6500~K effective temperature model as functions of pulsation phase. Peak values are the maximum value of a quantity throughout the 2D or 3D computational mesh at any given time. Note that these peak quantities are generally smaller for the 2D calculation than for the 3D calculation. 
}
\label{fig:6300-con-time-dep}
\end{figure}

\begin{figure}
\center
\plotone{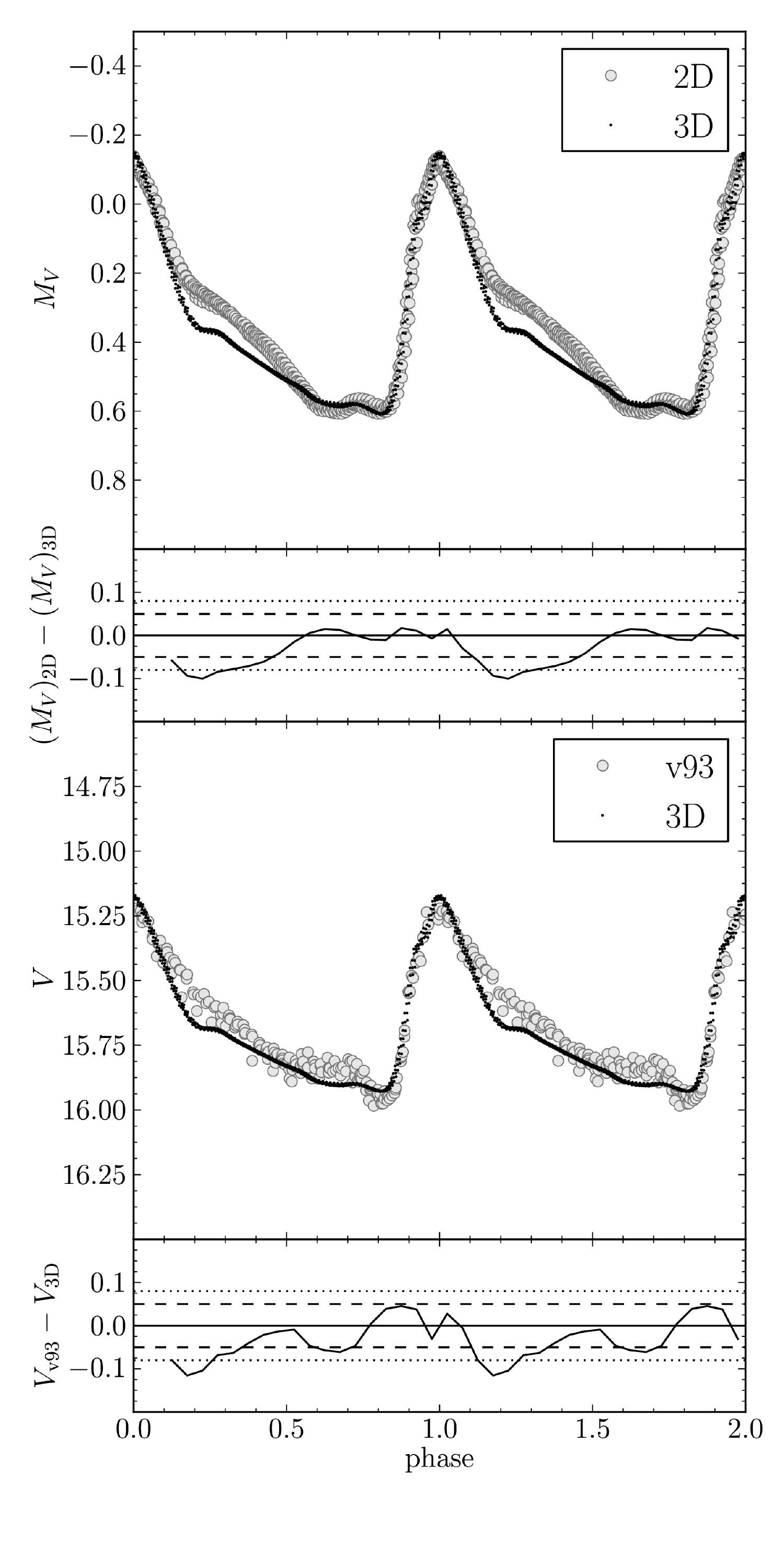}
\caption{The top panel shows the comparison of the 2D and 3D light curves for the 6500~K model. The bottom panel shows the 3D light curved compared to observations of v93 in M3 by \cite{Cacciari-2005}}
\label{fig:6500-light-curves}
\end{figure}

We have indicated that the time dependent behavior of convection as a function of pulsation phase is generally the same in 2D and 3D, based on general trends in the convective velocity and the warping of isothermal surfaces. Here we wish to examine the relative strength of convection a little more quantitatively in both types of calculations. This is not overly straightforward because we need a definition of the convective strength which can account for the differences in the flow patterns in 2D and 3D. This comparison will be made for a 6500~K model at full amplitude in both 2D and 3D. To proceed, we need to compute the convective flux for the two cases. There is no explicit expression for the convective flux in the conservation equations, only expressions for the total energy advection, the $PdV$ work, the conversion of the subgrid scale kinetic energy into heat, and the radiation terms \citep[see equation~8 of][]{Geroux-2013b}. Thus, the energy equation includes the energy balance of the pulsation, radiation, and convection without explicitly dividing the flow into that associated with pulsation and that associated with convection. However, we would like to examine the behavior of the (radial) convective flux, which we will approximate by
\begin{equation}
\label{eq:con-lum}
F_{{\rm conv.}}=c_P\rho \left(v_r-v_{r0}\right)\Delta T.
\end{equation}
where $v_r$ is the radial velocity of a given zone, $v_{r0}$ is the velocity of the coordinate system, and $\Delta T$ is the difference between the temperature in the zone from the horizontal average temperature at that radial zone. Recall that the velocity of the coordinate system is that required for the mass in the spherical shell to remain constant throughout the calculation \citep[see discussion in][]{Geroux-2011}. 

One possible comparison is between the maximum convective fluxes anywhere through the computational mesh at a particular time. Once convection has developed sufficiently, this will be in a downward moving column in either 2D or 3D. This maximum convective flux is merely the maximum value of the flux given in equation (1) over all the zones in the calculation. However, one could argue that the strength of convection should be measured by the amount of convective energy transport through a spherical surface. Here we must add up all the convective fluxes from all the zones at a given radius, with the individual zone surface areas taken into account. For convenience, we turn this into a convective luminosity. A comparison of these two convective flux related quantities in 2D and 3D will not necessarily yield the same result because the fraction of the surface area taken up by the downward moving material is quite different in the differing dimensions. Specifically, the 2D extension into 3D would have the downward convective flows moving in a long “trench” not shown in the 3D simulations.

We present the results of such a 2D -- 3D comparison in Figure~\ref{fig:6300-con-time-dep} for an effective temperature of 6500~K. The top panel shows the comparison of the 2D and 3D maximum convective flux, and the second panel shows the maximum convective luminosity computed as described above. The maximum convective flux is higher in 3D than 2D, while the opposite is true for the convective luminosity. Clearly, the fraction of the surface taken up by the large convective flux situated in the downward flow makes the difference. Having said that, we note that the relative time dependence of either the maximum convective flux or the convective luminosity is quite similar in 2D and 3D. The convective energy transport increases markedly during the latter phases of pulsational contraction and decreases during early expansion. As noted by \cite{Deupree-1977a} and also by \citep{Gastine-2011} this is the type of behavior which leads to a decrease in the pulsational driving by the ionization zones. This is not to say that this time dependent behavior is the sole property affecting pulsational growth or decay. For example, \cite{Gastine-2011} present an example in which the static model density stratification can appreciably affect the pulsation amplitude. While we do not believe this is an issue in this particular case because the pulsational kinetic energy is so much larger than the convective kinetic energy, we have not done a suite of calculations covering the possible range in the physical and model properties to determine if any of these affect the pulsation amplitudes.

The peak convective velocity, shown in the third panel of Figure~\ref{fig:6300-con-time-dep}, appears to be somewhat higher in 3D than 2D. The largest difference is at maximum convective flux, but the velocity differences are not that large. The maximum horizontal temperature variation also appears to be a little larger in 3D than 2D at maximum convective flux. The combination of these two differences are responsible for the increased maximum convective flux in 3D.

Of course, the crucial test in modelling RR Lyrae stars at full amplitude is the light curve. We compare the 2D light curve to the 3D light curve for the 6500~K full amplitude model and the 3D light curve to that of V93 in M3 \citep{Cacciari-2005} in Figure~\ref{fig:6500-light-curves}. We first note that the amplitude of the light curve and the rising light segments of all three light curves agree well. The 2D and 3D light curves differ in the rate of decline from maximum light and then in the new slope between phases 0.2 and 0.6. The 2D calculation actually agrees better with the observations during declining light, although the 3D slope between phases 0.2 and 0.6 is closer to that of V93 than is the 2D slope. The reasons for these differences are unknown, and more 3D calculations are required to determine the sensitivity of the light curves to parameters of the model and the zoning. The completion of the 3D models at other effective temperatures will indicate whether this is a global problem or confined to this one model.

\section{DISCUSSSION}
We have computed a number of 3D hydrodynamic models of RR Lyrae variables, one of which has now reached full amplitude. The convective flow pattern, of course, is genuinely 3D and thus different from that found in our previous 2D models \citep{Geroux-2013b}. However, the differences of the effects between 2D versus 3D convection on pulsation appear to be comparatively modest. The phase dependent behavior of the peak convective flux is quite similar between the 2D and 3D models, and the 3D models decrease the pulsational growth rate by only about 0.1\% per period compared to the 2D models. The comparison between light curves from the 2D and 3D calculations for the one 3D model at full amplitude are somewhat different during falling light, although the amplitude of the pulsation and the rising part of the light curve are quite similar. As full amplitude 3D calculations are completed, we should be able to determine how pervasive these differences are, particularly closer to the red edge. Also, very little has been done in terms of parameter studies in the 3D calculations. These remain difficult simply because 3D calculations to full amplitude require so much time.

The relatively small differences between the 2D and 3D calculations in terms of the effects of convection on pulsation should be considered good news. This suggests that the effects of different masses, luminosities, and compositions can probably be mapped out in 2D instead of the full 3D. Thus, the time can be shortened considerably because the 2D calculations take only days to weeks, whereas the 3D calculations require several months to reach full amplitude.

\acknowledgments
The authors gratefully acknowledge the support of ACEnet, both for providing high performance computing in Atlantic Canada and for an ACEnet Research Fellowship to CMG. We also thank ACEnet for the use of the Data Cave in visualizing the 3D calculations. As anyone who has performed significant 3D simulations knows, visualizing the results is almost as difficult as the calculations themselves, and the Data Cave animations made this possible. ACEnet is funded by the Canada Foundation for Innovation and provincial funding agencies of Nova Scotia, New Brunswick, and Newfoundland and Labrador. CMG received partial financial support during writing from a Consolidated STFC grant (ST/J001627/1).

\bibliographystyle{apj}

\end{document}